\documentclass[12pt]{article}
\usepackage{graphicx}
\usepackage{cite}
\usepackage{color}
\newcommand{\bm}[1]{\mbox{\boldmath $#1$}}
\newcommand{\fnd}[2]{\frac{\textstyle #1}{\textstyle #2}}

\newcommand{\abs}[1]{\left| #1\right|}
\newcommand{\fndrs}[4]{\fnd{\raisebox{#1}{$#2$}}{\raisebox{#3}{$#4$}}}

\newcommand{\dissum}[2]{\displaystyle \sum_{#1}^{#2}}

\begin{document}
\title{The spectrum of charmonium\\
in the Resonance-Spectrum Expansion}
\author{
Eef van Beveren$^{1}$ and George Rupp$^{2}$\\ [10pt]
{\small\it $^{1}$Centro de F\'{\i}sica Computacional,
Departamento de F\'{\i}sica,}\\
{\small\it Universidade de Coimbra, P-3004-516 Coimbra, Portugal}\\
{\small\it eef@teor.fis.uc.pt}\\ [10pt]
{\small\it $^{2}$Centro de F\'{\i}sica das Interac\c{c}\~{o}es Fundamentais,
Instituto Superior T\'{e}cnico,}\\
{\small\it Universidade T\'{e}cnica de Lisboa, Edif\'{\i}cio Ci\^{e}ncia,
P-1049-001 Lisboa, Portugal}\\
{\small\it george@ist.utl.pt}\\ [10pt]
{\small PACS number(s):
11.15.Tk,  
11.55.Jy,  
12.40.Yx,  
14.40.Gx  
}
}
\maketitle
\begin{abstract}
We argue that the resonance-like structures
$Y$(4260) \cite{PRL95p142001,PRL99p182004},
$Y$(4360), $Y$(4660) \cite{PRL99p142002}
and $Y$(4635) \cite{ARXIV08074458},
which were recently reported to have been observed in experiment,
are non-resonant manifestations of the Regge zeros
that appear in the production amplitude
of the Resonance-Spectrum Expansion.
Charmonium $c\bar{c}$ states are visible
on the slopes of these enhancements.
\end{abstract}

In the Resonance-Spectrum Expansion (RSE) \cite{IJTPGTNO11p179},
which is based on the model of Ref.~\cite{PRD21p772},
the meson-meson scattering amplitude is given by an expression
of the form (here restricted to the one-channel case)
\begin{equation}
\bm{T}(E)=\left\{ -2\lambda^{2}\mu p
j_{\ell}^{2}\left( pr_{0}\right)
\dissum{n=0}{\infty}
\fndrs{5pt}{\abs{g_{nL(\ell )}}^{2}}{-2pt}{E-E_{nL(\ell )}}
\right\}
\bm{\Pi}(E)
\;\;\; ,
\label{Tamplitude}
\end{equation}
where $p$ is the center-of-mass (CM) linear momentum,
$E=E(p)$ is the total invariant two-meson mass,
$j_{\ell}$ and $h^{(1)}_{\ell}$ are
the spherical Bessel function and Hankel function of the first kind,
respectively, $\mu$ is the reduced two-meson mass,
and $r_{0}$ is a parameter with dimension mass$^{-1}$,
which can be interpreted as the average string-breaking distance.
The coupling constants $g_{NL}$,
as well as the relation between $\ell$ and $L=L(\ell)$,
were determined in Ref.~\cite{ZPC21p291}.
The overall coupling constant $\lambda$,
which can be formulated in a flavor-independent manner,
represents the probability of quark-pair creation.
The dressed partial-wave RSE propagator
for strong interactions is given by
\begin{equation}
\bm{\Pi}_{\ell}(E)=\left\{
1-2i\lambda^{2}\mu pj_{\ell}\left( pr_{0}\right)
h^{(1)}_{\ell}\left( pr_{0}\right)
\dissum{n=0}{\infty}
\fndrs{5pt}{\abs{g_{NL}}^{2}}{-2pt}{E-E_{NL}}
\right\}^{-1}
\;\;\; .
\label{propagator}
\end{equation}

This propagator has the very intriguing property
that it vanishes for $E\to E_{NL}$.
We will show in the following that this phenomenon
can be, and has indeed been, observed in experiment,
but not in scattering processes,
as one easily verifies that
the RSE amplitude for strong scattering (\ref{Tamplitude})
does not vanish in the limit $E\to E_{NL}$.
However, for strong production processes, we
derived in Ref.~\cite{AP323p1215},
following a procedure similar to the one by Roca, Palomar, Oset, and
Chiang \cite{NPA744p127},
a relation between the production amplitude \bm{P}
and the scattering amplitude \bm{T}, reading
\begin{equation}
\bm{P}_{\ell}=
j_{\ell}\left( pr_{0}\right)
+i\,\bm{T}_{\ell}h^{(1)}_{\ell}\left( pr_{0}\right)
\;\;\; ,
\end{equation}
which, using Eqs.~(\ref{propagator}) and (\ref{Tamplitude}),
can also be written as
\begin{equation}
\bm{P}_{\ell}=
j_{\ell}\left( pr_{0}\right)
\bm{\Pi}_{\ell}(E)
\;\;\; .
\label{Production}
\end{equation}
From this expression we find,
by the use of Eq.~(\ref{propagator}),
that the production amplitude
tends to zero when $E\to E_{NL}$.
This effect must be visible in experimental
strong production cross sections.

Actually, in Ref.~\cite{AP323p1215}
we found, for the complete production amplitude
in the case of multi-channel processes,
that Eq.~(\ref{Production}) represents the leading term,
and that the remainder is expressed
in terms of the inelastic components of the scattering amplitude.
The latter terms do not vanish in the limit $E\to E_{NL}$,
as we have seen above.
Hence, the production amplitude only
vanishes {\it approximately} \/in this limit,
in case inelasticity is suppressed.

The reaction of electron-positron annihilation
into multi-hadron final states
takes basically place via one photon,
hence with $J^{PC}=1^{--}$ quantum numbers.
Consequently, when the photon materializes into
a pair of current quarks,
which couple via the $q\bar{q}$ propagator
to the final multi-hadron state,
we may assume that the intermediate propagator
carries the quantum numbers of the photon.
Moreover, alternative processes are suppressed.

We may thus conclude that,
if we want to discover
whether the propagator really vanishes at $E\to E_{NL}$,
then the ideal touchstone is
$e^{+}e^{-}$ annihilation into multi-hadron states.
Furthermore, there also exist predictions for the values of $E_{NL}$,
with $L=0$ or $L=2$,
given by the parameter set of Ref.~\cite{PRD27p1527}.
For $c\bar{c}$ one finds
in the latter paper
$E_{0,0}=3.409$ GeV and $\omega=0.19$ GeV,
which results for the higher
$c\bar{c}$ confinement states in the spectrum
$E_{1,0}=E_{0,2}=3.789$ GeV,
$E_{2,0}=E_{1,2}=4.169$ GeV,
$E_{3,0}=E_{2,2}=4.549$ GeV, \ldots .
\clearpage

\begin{figure}[htbp]
\begin{center}
\begin{tabular}{cc}
\includegraphics[width=200pt]{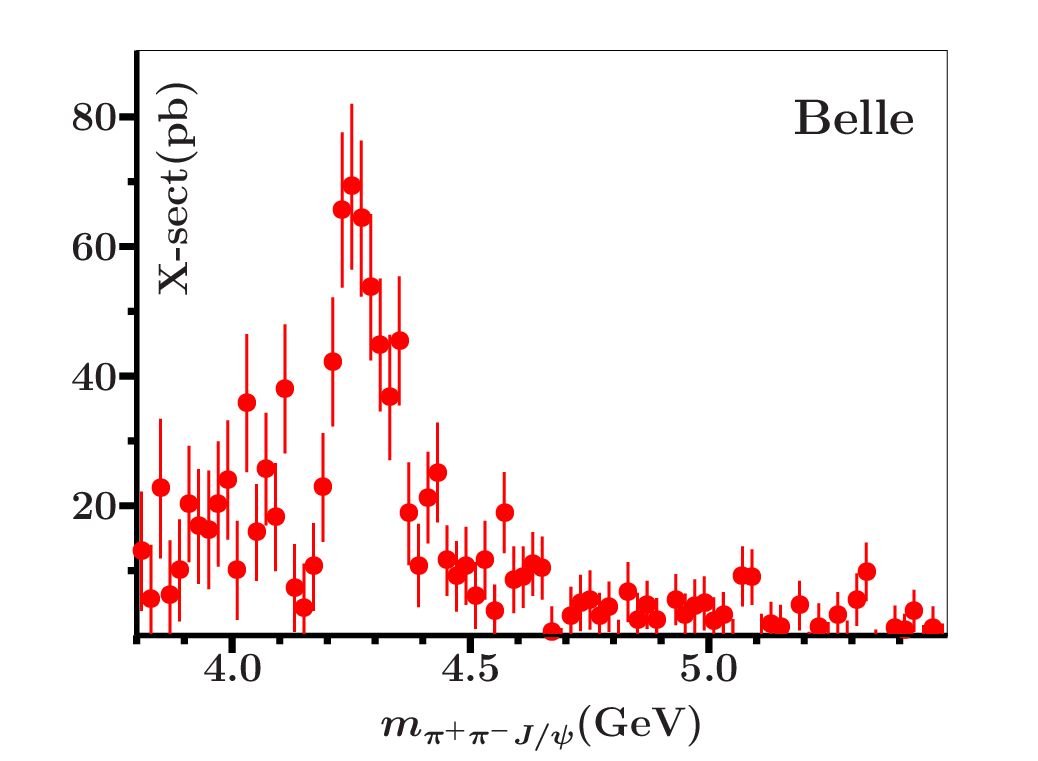} &
\includegraphics[width=200pt]{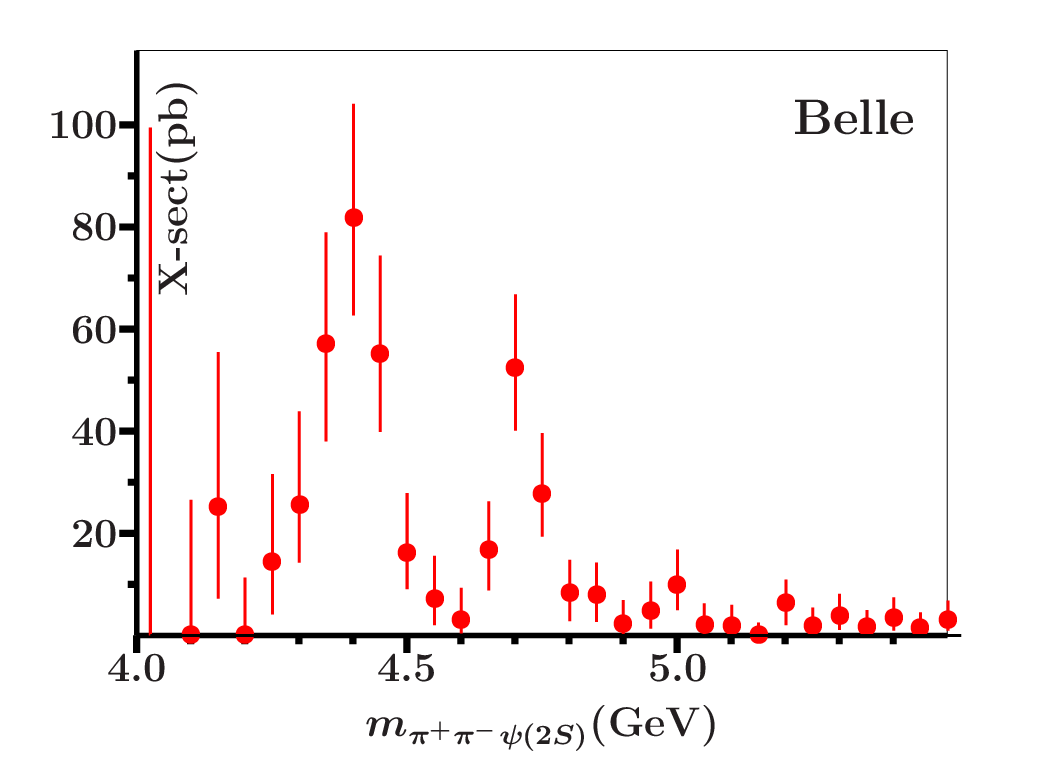}\\
(a) & (b)\\ [-20pt]
\end{tabular}
\end{center}
\caption{\small The $J/\psi\pi^{+}\pi^{-}$ (a)
and $\psi(2S)\pi^{+}\pi^{-}$ (b) invariant-mass distributions
for respectively the reactions $e^{-}e^{+}\to J/\psi\pi^{+}\pi^{-}$
and $e^{-}e^{+}\to\psi(2S)\pi^{+}\pi^{-}$.
The data, from the Belle Collaboration,
are taken from Ref.~\cite{PRL99p182004} (a)
and Ref.~\cite{PRL99p142002} (b).}
\label{Belle}
\end{figure}

The latter two levels of the $c\bar{c}$ confinement spectrum
can indeed be clearly observed in experiment.
For example, the non-resonant signal in
$e^{+}e^{-}\to\pi^{+}\pi^{-}\psi (2S)$ \cite{PRL99p142002},
which is depicted in Fig.~\ref{Belle}b,
is divided into two substructures
\cite{PRD77p014033,PRD78p014032,PLB665p26},
since the full $c\bar{c}$ propagator (\ref{propagator}),
dressed with meson loops,
vanishes at $E_{3}=4.55$ GeV
\cite{PRD27p1527}.
In the same set of data, one may observe a lower-lying zero
at $E_{2}=4.17$ GeV
\cite{PRD27p1527},
more distinctly visible in the data on
$e^{+}e^{-}\to\pi^{+}\pi^{-}J/\psi$ \cite{PRL99p182004},
which are depicted in Fig.~\ref{Belle}a.
The true $c\bar{c}$ resonances
can be found on the slopes
of the above-mentioned non-resonant structures,
unfortunately with little statistical significance,
if any.

\begin{figure}[htbp]
\begin{center}
\begin{tabular}{cc}
\includegraphics[height=150pt, angle=0]{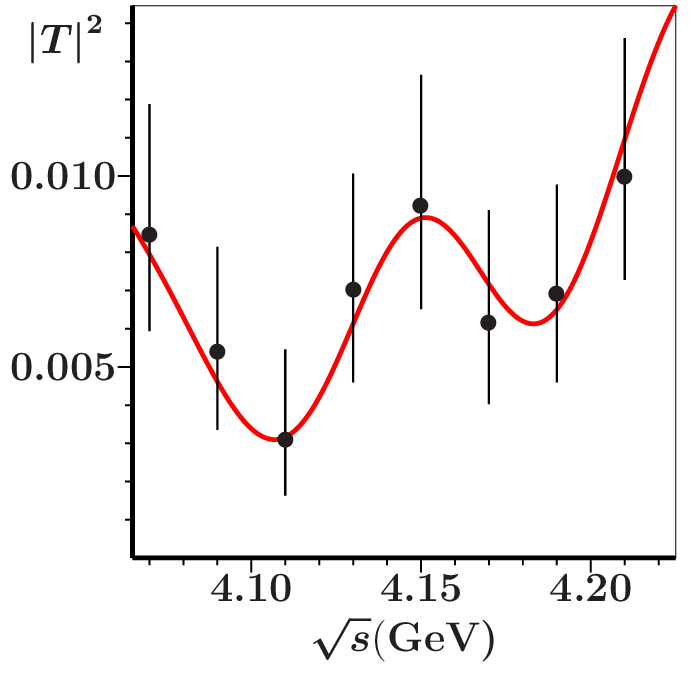}
&
\includegraphics[height=150pt, angle=0]{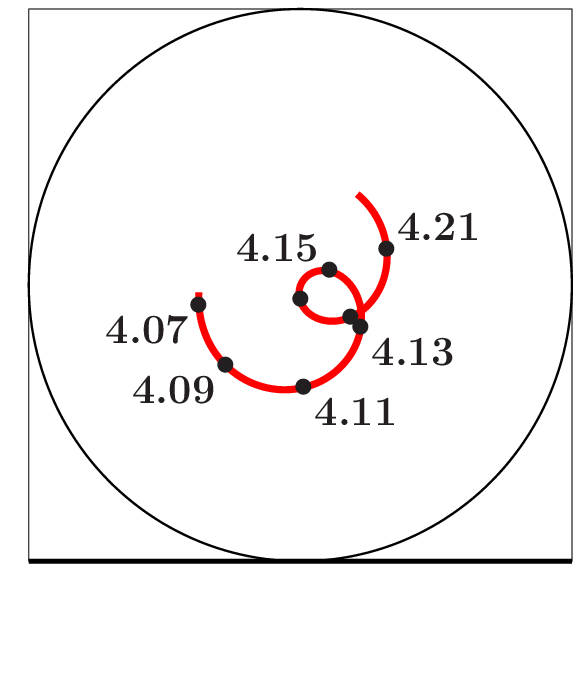} \\[-8mm]
\end{tabular}
\end{center}
\caption[]{Amplitudes and phases for $\psi (2D)$ at 4160 MeV.
The data are taken from Ref.~\cite{PRL95p142001} with arbitrary normalization.}
\label{psi2D}
\end{figure}

In fact, in Ref.~\cite{PRL95p142001},
where the BaBar Collaboration announced the observation of
the $Y(4260)$ structure in $e^{+}e^{-}\to\pi^{+}\pi^{-}J/\psi$,
one reads:
``{\it no other structures are evident at the masses
of the quantum number} $J^{PC}=1^{--}$ {\it charmonium states,
i.e., the} $\psi(4040)$, $\psi(4160)$, {\it and} $\psi(4415)$''.
However, in Ref.~\cite{HEPPH0605317}, we demonstrated that the
BaBar data at about 4.15 GeV are consistent with
the mass and width of the $\psi(4160)$
(see Fig.~\ref{psi2D}).
Here, we will show that also the $\psi(4415)$
is clearly visible in the BaBar data,
possibly even with enough statistical significance.
\clearpage

So we indeed observe minima in production processes,
which confirm vanishing $q\bar{q}$ propagators.
Moreover, the $q\bar{q}$ confinement spectrum predicted
25 years ago in Ref.~\cite{PRD27p1527}
seems to agree well with experimental observations
for vector mesons.
Accordingly, we expect vector-meson $q\bar{q}$ resonances
associated with each of the Regge states:
one ground state, dominantly in a $q\bar{q}$ $S$-wave,
and two resonances for each of the higher excited Regge states, viz.\
one dominantly in an $S$-wave, and the other mostly in a $D$-wave.
In Ref.~\cite{ARXIV08091151}, we reported on indications for
four, possibly five, new $c\bar{c}$ vector states,
in the $e^+e^-\to\Lambda_c^+\Lambda_{c}^-$ amplitudes
of the Belle Collaboration \cite{ARXIV08074458}.
Here, we will just concentrate on the $\psi(4S)$.

\begin{figure}[htbp]
\begin{center}
\begin{tabular}{c}
\includegraphics[width=443pt]{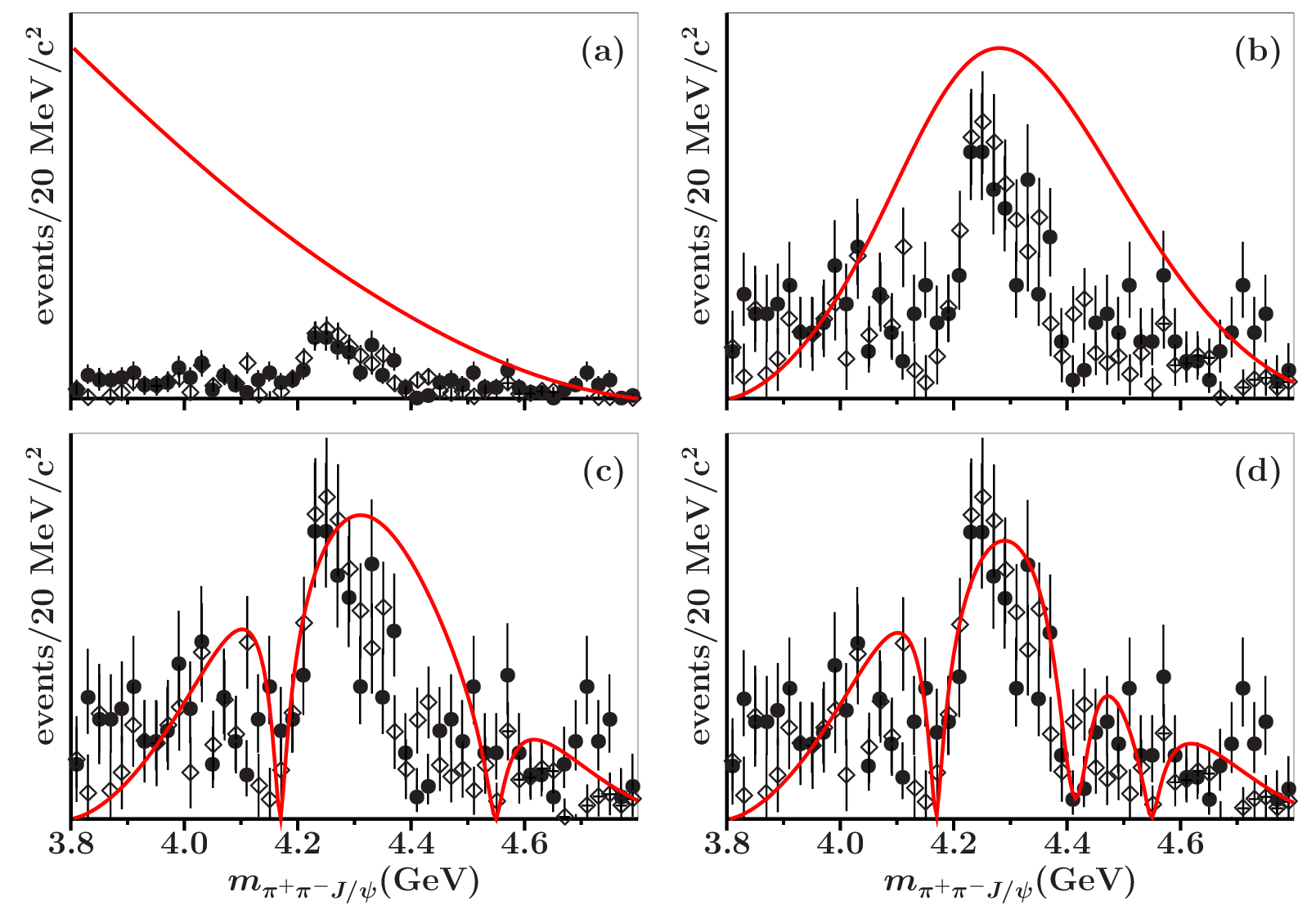}\\ [-20pt]
\end{tabular}
\end{center}
\caption{\small The $J/\psi(\pi^{+}\pi^{-})$ invariant-mass distributions
for the reaction $e^{-}e^{+}\to J/\psi\pi^{+}\pi^{-}$.
Data are taken from Ref.~\cite{PRL95p142001} ($\bullet$)
and Ref.~\cite{PRL99p182004} ($\diamond$).
The theoretical results (solid lines) are also discussed in the text:
(a) shows the distribution for a non-resonant structureless
$c\bar{c}$ propagator;
(b) and (c) show the effect of Regge zeros in the $c\bar{c}$
vector propagator, thereby surpressing the contributions
of its $c\bar{c}$ poles;
(b) for just the Regge zero at 3.79 GeV;
(c) for the zeros at 4.17 and 4.55 GeV as well;
(d) shows the additional effect of the $c\bar{c}$ resonance pole
in the propagator at $4.415-i0.036$ GeV \cite{PLB667p1}.}
\label{y4260ac}
\end{figure}

A full description of the $\pi^{+}\pi^{-}J/\psi$
involves a three-body calculation.
In the present work, however, we will limit us to an effective
two-body calculation for $\left(\pi^{+}\pi^{-}\right) J/\psi$,
assuming for the $\pi^{+}\pi^{-}$ effective mass just a fraction of
the available phase space.
Furthermore, we assume an $S$-wave for the relative orbital angular
momentum of the $\pi^{+}\pi^{-}$-$J/\psi$ system.
Under these assumptions, we obtain for the amplitude
the result depicted in Fig.~\ref{y4260ac}a,
for the case that the propagator of Eq.~(\ref{propagator}) is substituted
by a structureless vertex.

As expected, we observe no resonances
in the amplitude of Fig.~\ref{y4260ac}a.
Next, we suppress the effect of the resonance poles
in the propagator of Eq.~(\ref{propagator}),
such that the zeros at $E=E_{NL}$
dominate production (see Eq.~(\ref{Production})).
The resulting amplitudes are shown in Figs.~\ref{y4260ac}b
and \ref{y4260ac}c.
We observe that now our theoretical amplitude
is in rather good agreement with the data.
There is an excess of data for energies below 4.0 GeV,
stemming from the tail of the $\psi(3685)$ resonance,
which dominates the amplitude at lower masses
and which is not accounted for in our amplitude.
Furthermore, in Ref.~\cite{HEPPH0605317}
we discussed the $\psi(4160)$ resonance,
which, since not accounted for here, leads to an overestimate
of the BaBar data by our theoretical amplitude.

However, there is a rather large overestimate visible
in Fig.~\ref{y4260ac}c at the mass of the $\psi(4415)$.
In Fig.~\ref{y4260ac}d, we show that the difference between data
and our non-resonant amplitude
can be perfectly explained by accounting for a $c\bar{c}$
resonance with mass and width consistent with the $\psi(4415)$.
Moreover, the experimental error bars indicate that
sufficient statistics is available to include this resonance in
a data analysis for the non-resonant $Y$ structures.
\vspace{10pt}

{\bf Summarizing},
we have shown that the $c\bar{c}$ confinement spectrum,
which underlies scattering and production of multi-meson systems
containing charmonium $q\bar{q}$ pairs,
can be observed in production amplitudes.
Moremore, we have shown that the $c\bar{c}$ resonance poles
are present in the $e^{-}e^{+}\to J/\psi\pi^{+}\pi^{-}$ amplitude.
\vspace{10pt}

We wish to thank Xiang Liu for very fruitful discussions.
This work was supported in part by the {\it Funda\c{c}\~{a}o para a
Ci\^{e}ncia e a Tecnologia} \/of the {\it Minist\'{e}rio da Ci\^{e}ncia,
Tecnologia e Ensino Superior} \/of Portugal, under contract
POCI/\-FP/\-81913/\-2007.

\newcommand{\pubprt}[4]{#1 {\bf #2}, #3 (#4)}
\newcommand{\ertbid}[4]{[Erratum-ibid.~#1 {\bf #2}, #3 (#4)]}
\def\AP{Ann.\ Phys.}
\def\IJTPGTNO{Int.\ J.\ Theor.\ Phys.\ Group Theor.\ Nonlin.\ Opt.}
\def\NPA{Nucl.\ Phys.\ A}
\def\PLB{Phys.\ Lett.\ B}
\def\PRD{Phys.\ Rev.\ D}
\def\PRL{Phys.\ Rev.\ Lett.}
\def\ZPC{Z.\ Phys.\ C}


\begin{thebibliography}{16}
\bibitem{PRL95p142001}
B.~Aubert {\it et al.}  [BABAR Collaboration],
{\it Observation of a broad structure in the $\pi^{+}\pi^{-}J/\psi$
mass spectrum around 4.26-GeV/c$^{2}$},
\pubprt{\PRL}{95}{142001}{2005}
[arXiv:hep-ex/0506081].

\bibitem{PRL99p182004}
C.~Z.~Yuan {\it et al.}  [Belle Collaboration],
{\it Measurement of $e^{+}e^{-}\to\pi^{+}\pi^{-}J/\psi$ cross section
via initial-state radiation at Belle},
\pubprt{\PRL}{99}{182004}{2007}
[arXiv:0707.2541 [hep-ex]].

\bibitem{PRL99p142002}
X.~L.~Wang {\it et al.}  [Belle Collaboration],
{\it Observation of two resonant structures in
$e^{+}e^{-}\to\pi^{+}\pi^{-}\psi (2S)$ via
initial-state radiation at Belle},
\pubprt{\PRL}{99}{142002}{2007}
[arXiv:0707.3699 [hep-ex]].

\bibitem{ARXIV08074458}
G.~Pakhlova {\it et al.}  [Belle Collaboration],
{\it Observation of a near-threshold enhancement in the
$e^{+}e^{-}\to\Lambda_{c}^{+}\Lambda_{c}^{-}$
cross section using initial-state radiation},
arXiv:0807.4458 [hep-ex].

\bibitem{IJTPGTNO11p179}
E.~van Beveren and G.~Rupp,
{\it Reconciling the light scalar mesons with Breit-Wigner resonances as
well as the quark model},
\pubprt{\IJTPGTNO}{11}{179}{2006}
[arXiv:hep-ph/0304105].

\bibitem{PRD21p772}
E.~van Beveren, C.~Dullemond, and G.~Rupp,
{\it Spectra and strong decays of $c\bar{c}$ and $b\bar{b}$ states},
\pubprt{\PRD}{21}{772}{1980}
\ertbid{\ D}{22}{787}{1980}.

\bibitem{ZPC21p291}
E.~van Beveren,
{\it Coupling constants and transition potentials for hadronic decay
modes of a meson},
\pubprt{\ZPC}{21}{291}{1984}
[arXiv:hep-ph/0602246].

\bibitem{AP323p1215}
E.~van Beveren and G.~Rupp,
{\it Relating multichannel scattering and production amplitudes
in a microscopic OZI-based model},
\pubprt{\AP}{323}{1215}{2008}
[arXiv:0706.4119].

\bibitem{NPA744p127}
L.~Roca, J.~E.~Palomar, E.~Oset and H.~C.~Chiang,
{\it Unitary chiral dynamics in $J/\psi\to V P P$ decays
and the role of scalar mesons},
\pubprt{\NPA}{744}{127}{2004}
[arXiv:hep-ph/0405228].

\bibitem{PRD27p1527}
E.~van Beveren, G.~Rupp, T.~A.~Rij\-ken, and C.~Dullemond,
{\it Radial spectra and hadronic decay widths of light and heavy mesons},
\pubprt{\PRD}{27}{1527}{1983}.

\bibitem{PRD77p014033}
G.~J.~Ding, J.~J.~Zhu and M.~L.~Yan,
{\it Canonical charmonium interpretation for Y(4360) and Y(4660)},
\pubprt{\PRD}{77}{014033}{2008}
[arXiv:0708.3712 [hep-ph]].

\bibitem{PRD78p014032}
Z.~Q.~Liu, X.~S.~Qin and C.~Z.~Yuan,
{\it Combined fit to BaBar and Belle data on
$e^{+}e^{-}\to\pi^{+}\pi^{-}\psi (2S)$},
\pubprt{\PRD}{78}{014032}{2008}
[arXiv:0805.3560 [hep-ex]].

\bibitem{PLB665p26}
F.~K.~Guo, C.~Hanhart and U.~G.~Meissner,
{\it  Evidence that the $Y$(4660) is a $f_{0}$(980)$\psi '$ bound state},
\pubprt{\PLB}{665}{26}{2008}
[arXiv:0803.1392 [hep-ph]].

\bibitem{HEPPH0605317}
E.~van Beveren and G.~Rupp,
{\it Is the Y(4260) just a coupled-channel signal?},
arXiv:hep-ph/0605317.

\bibitem{ARXIV08091151}
E.~van~Beveren, X.~Liu, R.~Coimbra, and G.~Rupp,
{\it Possible $\psi(5S)$, $\psi(4D)$, $\psi(6S)$, and
$\psi(5D)$ signals in $\Lambda_{c}\bar{\Lambda}_{c}$},
arXiv:0809.1151 [hep-ph].

\bibitem{PLB667p1}
C.~Amsler {\it et al.} \/[Particle Data Group Collaboration],
{\it Review of Particle Physics},
\pubprt{\PLB}{667}{1}{2008}.
\end{thebibliography}
\end{document}